\documentclass[aps,prb,showpacs,twocolumn,superscriptaddress]{revtex4-1}

\usepackage{amsmath}
\usepackage{amsfonts}
\usepackage{amssymb}
\usepackage{dsfont}
\usepackage{epsfig}
\usepackage{epstopdf}
\usepackage{physics}
\usepackage{microtype}
\usepackage{bm}

\usepackage{graphicx}

\usepackage{hyphenat}
\usepackage{hyperref}
\usepackage{placeins}	
\usepackage{tikz}
\usepackage{xcolor}
\usepackage{color}
\usepackage{xspace}
\usepackage{accents}
\usepackage{braket}
\usepackage{subcaption}
\usepackage{graphicx}  
\usepackage{todonotes}
\usepackage{comment}
\usepackage{ragged2e}
\usepackage{soul}
\DeclareCaptionJustification{justified}{\justifying}
\usepackage{array}
\newcolumntype{C}[1]{>{\centering\arraybackslash}m{#1}}
\usepackage{overpic}

\usepackage[normalem]{ulem}

\usepackage{booktabs}
\usepackage{diagbox}
\usepackage{siunitx}
\AtBeginDocument{
	\heavyrulewidth=.08em
	\lightrulewidth=.05em
	\cmidrulewidth=.03em
	\belowrulesep=.65ex
	\belowbottomsep=0pt
	\aboverulesep=.4ex
	\abovetopsep=0pt
	\cmidrulesep=\doublerulesep
	\cmidrulekern=.5em
	\defaultaddspace=.5em
}
\bibpunct{[}{]}{;}{n}{}{,~}

\begin{document}
	
\title{Quantum criticality and excitations of a long-range anisotropic XY chain in a transverse field}
\author{Patrick Adelhardt}
\affiliation{Lehrstuhl f\"ur Theoretische Physik I, Staudtstra{\ss}e 7, FAU Erlangen-N\"urnberg, D-91058 Erlangen, Germany}
\author{Jan Alexander Koziol}
\affiliation{Lehrstuhl f\"ur Theoretische Physik I, Staudtstra{\ss}e 7, FAU Erlangen-N\"urnberg, D-91058 Erlangen, Germany}
\author{Andreas Schellenberger}
\affiliation{Lehrstuhl f\"ur Theoretische Physik I, Staudtstra{\ss}e 7, FAU Erlangen-N\"urnberg, D-91058 Erlangen, Germany}
\author{Kai Phillip Schmidt}
\affiliation{Lehrstuhl f\"ur Theoretische Physik I, Staudtstra{\ss}e 7, FAU Erlangen-N\"urnberg, D-91058 Erlangen, Germany}

\begin{abstract}
	The critical breakdown of a one-dimensional quantum magnet with long-range interactions is studied by investigating the high-field polarized phase of the anisotropic XY model in a transverse field for the ferro- and antiferromagnetic case. While for the limiting case of the isotropic long-range XY model we can extract the elementary one quasi-particle dispersion analytically and calculate two quasi-particle excitation energies quantitatively in a numerical fashion, for the long-range Ising limit as well as in the intermediate regime we use perturbative continuous unitary transformations on white graphs in combination with classical Monte Carlo simulations for the graph embedding to extract high-order series expansions in the thermodynamic limit. This enables us to determine the quantum-critical breakdown of the high-field polarized phase by analyzing the gap-closing including associated critical exponents and multiplicative logarithmic corrections. In addition, for the ferromagnetic isotropic XY model we determined the critical exponents $z$ and $\nu$ analytically by a bosonic quantum-field theory.
\end{abstract}

\maketitle

\section{Introduction} 
Quantum many-body systems are known to display a variety of fascinating physical properties ranging from exotic ground-states with unconventional properties like superconductivity or long-range entangled topological order to elementary excitations with fractional quantum numbers and non-standard particle statistics. Continuous quantum phase transitions between distinct phases of quantum matter can typically be understood in terms of a gap-closing of the relevant excitations displaying universal quantum criticality characterized by critical exponents. In the past most research has focused on correlated lattice models with short-range interactions, most prominently the nearest-neighbor transverse-field Ising model \cite{Pfeuty} or the XY model \cite{LiebSchulzMattis, McCoy, Barouch, Jimbo, Santos, Ray, Damle}. In the absence of geometric frustration, like for ferromagnetic interactions, the criticality of such systems is governed by the spatial dimension and the underlying symmetry group, while frustration gives rise to richer phase diagrams \cite{Kano_1953, Villain_1980, Shender_1982, Blankschtein1984, Moessner2001, Moessner2003, Powalski2013, Hermele_2004, Roechner2016}.
\par
In recent years correlated quantum systems with long-range interactions have been studied more intensively \cite{Bitko, Chakraborty, Mahmoudian, Mengotti, Lahaye, Peter, Bramwell, Castelnovo, Britton, Islam, Jurcevic, Richerme,  Bohnet, YangJiang, Fisher1972, Dutta, Knap, Suzuki1, Suzuki2, Yamazaki, Picco, Blanchard, Grassberger, Sak, Defenu2017, Zhangqi, Gori, Luijten, Angelini, Horita, Paulos, Behan1, Behan2, Behan3, Peter, Koffel, Sun, FeySchmidt1D, FeySchmidt2D, KoziolCylinders, Humeniuk, Saadatmand}. On the one hand, these systems are known to be experimentally relevant, e.g.~, in spin-ice materials in condensed matter physics \cite{Bramwell, Castelnovo}, or in quantum optics with cold ions \cite{Britton, Islam, Jurcevic, Richerme, Bohnet, YangJiang} or Rydberg atoms \cite{Labuhn}. On the other hand, these systems are interesting from a theoretical perspective, since they are known to behave differently compared to their short-range counterparts, e.g.~, there exist continuously varying critical exponents which can be understood in terms of an effective increase in the spatial dimension due to the long-range interaction \cite{Defenu_2020}. 
\par
Theoretically, the treatment of quantum many-body systems is notoriously complicated so that many investigations have mostly focused on one-dimensional paradigmatic and unfrustrated models like the transverse-field Ising chain or the Heisenberg chain with long-range dipolar interactions, which are still accessible by numerical techniques like exact diagonalization \cite{Knap, Sandvik2010, Zhangqi}, tensor network approaches \cite{Koffel, Sun, Vanderstraeten2018, Zhangqi}, quantum Monte Carlo simulations \cite{Sandvik2003, Baez2019, Yang2020, Humeniuk, Saadatmand}, or linked-cluster expansions \cite{FeySchmidt1D}. The latter two methods have been extended to frustrated two-dimensional systems only recently \cite{Sandvik2003, Baez2019, Yang2020, Humeniuk, Saadatmand, FeySchmidt2D, KoziolCylinders}.  
\par
The quantum-critical behavior of the unfrustrated long-range transverse-field Ising chain with ferromagnetic interactions is known to display three distinct regimes including the nearest-neighbor (1+1)D Ising universality class for short-range interactions, mean-field behavior for strong long-range interactions, and continuously varying critical exponents in between \cite{Sak, Dutta, Defenu2017, Defenu_2020, Behan1, Behan2}. In contrast, for antiferromagnetic interactions, the Ising exchange at odd and even distances compete so that numerical investigations by the density matrix renormalization group \cite{Sun} and linked-cluster expansions \cite{FeySchmidt1D, FeySchmidt2D} point towards a (1+1)D Ising universality for any strength of algebraically decaying interactions. The latter finding is obtained in linked-cluster expansions by analyzing the one-particle (dressed spin flip) gap-closing within the high-field polarized phase. Here we extend these calculations in two directions strengthening the above scenario. First, we consider the isotropic XY case which is to a great extent analytically tractable, since first-order perturbation theory becomes exact for the high-field gap. This allows us to calculate critical exponents of the high-field gap as well as two-particle excitations quantitatively even in the presence of long-range interactions. Second, we study a generalized long-range spin chain which interpolates between pure Ising and XY interactions. We find that the high-order series expansion becomes increasingly well behaved when tuning towards the isotropic XY model. As a consequence, for antiferromagnetic interactions, extrapolations converge better when deforming away from the pure Ising case. In contrast, for ferromagnetic interactions, the extrapolants exhibits best convergence in the intermediate interpolation regime.
\par
The article is organized as follows. In Sect.~\ref{sec::Model} we introduce the microscopic model. In Sect.~\ref{sec::Approach} we explain the technical aspects including the method of perturbative continuous unitary transformations (pCUT) as a linked-cluster expansion (LCE), the embedding procedure, and the extrapolation schemes employed. In Sect.~\ref{sec::IXY} we present analytical and numerical results for the limiting case of isotropic interactions. We proceed in Sect.~\ref{sec::AIXY} with the discussion of the results of the entire model. Finally, we draw conclusions in Sect.~\ref{sec::Conclusion}.

\section{Model} 
\label{sec::Model}
We consider the anisotropic XY chain with long-range interactions in a transverse field which is given by
%
\begin{equation}
\mathcal{H} =  h\sum_{i}  \sigma_{i}^{z} - \sum_{i,\delta>0}\frac{J_\delta}{2}\left[(1+\beta)\sigma_{i}^{x}\sigma_{i+\delta}^{x} + (1-\beta)\sigma_{i}^{y}\sigma_{i+\delta}^{y}\right]\, ,
\label{Eq::XY_Ham}
\end{equation}
%
where the Pauli matrices $\sigma^\kappa_{i}$ with $\kappa\in\{{x,y,z}\}$ describe spin-1/2 operators on the $i$-th lattice site of the chain, the transverse field strength $h>0$, and the coupling parameter $J_\delta=J \delta^{-\alpha}$ inducing ferromagnetic (antiferromagnetic) interactions for positive (negative) values. We confine the interpolation parameter $\beta\in [0,1]$ to tune the system from the XY chain with isotropic interactions ($\beta=0$) to  pure Ising-type interactions ($\beta=1$). The parameter $\alpha$ changes the long-range behavior from the limiting cases of an all-to-all coupling for $\alpha=0$ to a nearest-neighbor coupling for $\alpha=\infty$.
\par
In the limiting case of a vanishing coupling strength $J=0$ the system shows a non-degenerate $z$-polarized ground-state $\ket{\downarrow\dots\downarrow}$ while for a vanishing magnetic field $h=0$ the system exhibits magnetic ordering, either ferro- or antiferromagnetic depending on the sign of $J$, in $x$-direction ($y$-direction) for $\beta>0$ ($\beta<0$). When tuning the interpolation parameter $\beta$ from pure Ising-type interactions to isotropic XY interactions, ordering in the $x$-direction starts to compete with ordering in the $y$-direction. At any $\beta\neq0$ the system with nearest-neighbor interactions undergoes the same second-order phase transition at $J=\pm h$ belonging to the (1+1)dimensional [(1+1)D] Ising universality class from the paramagnetic phase to the (anti-)ferromagnetically ordered phase which breaks the $\mathbb{Z}_2$-symmetry of the Hamiltonian. At $\beta=0$ and $|J|>h$ the system displays an (Ising)$^{\text{2}}$ transition between $x$- and $y$-ordered phases as a function of $\beta$ \cite{Damle}. Furthermore, the system has multicritical points at $\beta=0$ and $J=\pm h$ where the critical lines of the two transitions meet. Though the phase diagram of the nearest-neighbor model has been studied extensively in the past \cite{LiebSchulzMattis, McCoy, Barouch, Jimbo, Santos, Ray, Damle}, its long-range counterpart is not well investigated except for the limiting cases of isotropic XY interactions \cite{Pazmandi1993, deSousa2005} and pure Ising-type interactions \cite{Fisher1972, Dutta, Knap, Suzuki1, Suzuki2, Yamazaki, Picco, Blanchard, Grassberger, Sak, Defenu2017, Zhangqi, Gori, Luijten, Angelini, Horita, Paulos, Behan1, Behan2, Behan3, Peter, Koffel, Sun, FeySchmidt1D, FeySchmidt2D, KoziolCylinders}. In the limit of the long-range transverse-field Ising model (LRTFIM) with antiferromagnetic interactions recent investigations \cite{Sun, FeySchmidt1D, FeySchmidt2D} indicate a (1+1)D Ising phase transition at any $\alpha$ like for the nearest-neighbor model. In contrast, for ferromagnetic interactions the system shows such critical behavior only for large values of $\alpha$ but falls into mean-field universality for small $\alpha$ and continuously varying exponents in between \cite{Sak, Dutta, Defenu2017, Defenu_2020, Behan1, Behan2}.
\par
In the following, we will perform high-order series expansions about the high-field limit $h\gg J$ where the fully polarized ground-state $\ket{\downarrow\dots\downarrow}$ serves as the unperturbed reference state $\ket{\text{ref}}$ and elementary excitations are local spin flips $\ket{\uparrow_j}$ at an arbitrary site $j$. We rescale the global energy spectrum of the Hamiltonian by $1/2h$ such that the excitation energy of a local spin flip is $1$. Further, we employ the Matsubara-Matsuda transformation \cite{Matsubara} to express the Hamiltonian exactly in terms of hard-core boson annihilation (creation) operators $\hat{b}_{i}^{(\dagger)}$, yielding
\begin{align}
\mathcal{H} = E_0+\mathcal{Q} - \sum_{i,\delta>0}&\frac{\lambda_{\delta}}{2}\left[(1+\beta)(\hat{b}_{i}^{\dagger}\hat{b}_{i+\delta}^{\phantom{\dagger}} + \hat{b}_{i}^{\dagger}\hat{b}_{i+\delta}^{\dagger})\right. \nonumber\\
+ (1&-\beta)\left.(\hat{b}_{i}^{\dagger}\hat{b}_{i+\delta}^{\phantom{\dagger}} - \hat{b}_{i}^{\dagger}\hat{b}_{i+\delta}^{\dagger}) + h.c.\right]\, ,
\label{Eq::XY_Ham_Op}
\end{align}
where we introduced $\lambda_{\delta}=J_\delta/2h$, the bare ground-state energy $E_0=-N/2$ and \mbox{$\mathcal{Q}=\sum_i \hat{b}^{\dagger}_{i}\hat{b}^{\phantom{\dagger}}_{i}$}, which counts the number of quasi-particles (QPs). These QPs of the high-field polarized phase are dressed spin flips, i.e., local spin flips dressed by quantum fluctuations induced by the anisotropic XY interactions.

\section{Approach}\label{sec::Approach}

In order to be self contained this section serves as a structured summary of the most important steps of the recent methodological extensions to the pCUT method to long-range quantum spin systems. Technically, we apply the pCUT method \cite{knetter2000perturbation, knetter2003structure} with the help of white graphs \cite{coester2015optimizing} along the same lines as done for the LRTFIM  \cite{FeySchmidt1D,FeySchmidt2D,KoziolCylinders} as well as for Heisenberg quantum spin models in the presence of quenched disorder \cite{hormann2018dynamic, hormann2020dynamic}. In particular, here we extended the white-graph expansion \cite{coester2015optimizing} to multiple couplings acting on the same bond. However, the understanding of this section is not essential for the discussion of the physical results in Sect. \ref{sec::IXY} and Sect. \ref{sec::AIXY}.

\subsection{pCUT method}\label{ssec::PCUT}
For the pCUT method we rewrite the Hamiltonian of Eq.~\eqref{Eq::XY_Ham_Op} as
\begin{align}
\mathcal{H} &= \mathcal{H}_0 + \mathcal{V} \nonumber\\
&= E_0 + \mathcal{Q} + \sum_{\delta > 0}^{\infty}\lambda_{\delta} \mathcal{V}(\delta)\, ,
\end{align}
where we have chosen the unperturbed part $\mathcal{H}_0$ to be the magnetic field term. The perturbation $\mathcal{V}$ decomposes into a sum of infinitely many perturbations associated with expansion parameters $\lambda_{\delta}$ depending on the distance $\delta$ of interacting spin pairs. As another prerequisite of the pCUT method, the perturbation $\mathcal{V}$ must be expressible as a sum 
\begin{equation}
\mathcal{V} = \sum_{m=-N}^{N}\hat{T}_m =\sum_{m=-N}^{N}\sum_{l}\hat{\tau}_{m,l}
\end{equation}
of operators $\hat{T}_{m}$ made up of local processes $\hat{\tau}_{m,l}$ on links $l$ connecting different sites on the chain. Such an operator changes the energy of the system by $m$ quanta satisfying the commutation relation $[\mathcal{Q},\hat{T}_m]=m\hat{T}_m$. In the case of Eq.~\eqref{Eq::XY_Ham_Op}, the perturbation reads as
\begin{equation}
\mathcal{V}=\hat{T}_{-2} + \hat{T}_{0} + \hat{T}_{2}\, .
\label{Eq::T_ops}
\end{equation}
\par
The conceptional idea of the pCUT method is to unitarily transform the original Hamiltonian, order by order in perturbation, to an effective, quasi-particle-conserving Hamiltonian $\mathcal{H}_{\text{eff}}$ which essentially reduces the initial many-body to an effective few-body problem. The effective pCUT Hamiltonian is then generally given by 
\begin{align}
\mathcal{H}_{\text{eff}} &= \mathcal{H}_0 + \sum_{\sum_j n_j = k}^{\infty} \lambda_{1}^{n_1} \dots \lambda_{k}^{n_k} \nonumber\\
&\times \sum_{\substack{\dim(\bm{m})=k, \\ \sum_i m_i=0}} C(\bm{m})\;\hat{T}_{m_1}\dots \hat{T}_{m_{k}}\, 
\label{Eq::H_eff}
\end{align}
with the exact coefficients $C(\bm{m})\in \mathbb{Q}$ and the condition $\sum_i m_i=0$ which reflects the quasi-particle conservation $[\mathcal{Q}, \mathcal{H}_{\text{eff}}]=0$. The effective Hamiltonian in Eq.~\eqref{Eq::H_eff} is independent of the specific form of the original Hamiltonian as long as the general prerequisites of the pCUT method are fulfilled. However, this model-independent part comes with a second model-dependent extraction process to bring $\mathcal{H}_{\text{eff}}$ in a normal-ordered form. The latter is done in an optimal fashion via a full graph decomposition which we briefly describe next.

\subsection{Graph decomposition}
\label{ssec:Graph_Decomp}
The full graph decomposition within the pCUT approach corresponds to a linked-cluster expansion. Indeed, since Eq.~\eqref{Eq::H_eff} is a cluster-additive quantity, we can exploit the linked-cluster theorem and rewrite the effective Hamiltonian as
\begin{align}
\mathcal{H}_{\text{eff}} = \mathcal{H}_0 &+ \sum_{\sum_j n_j = k}^{\infty} \lambda_{1}^{n_1} \dots \lambda_{k}^{n_k} \sum_{\substack{\dim(\bm{m})=k, \\ \sum_i m_i=0}}\sum_{\mathcal{G}_k} C(\bm{m}) \nonumber\\
&\times  \sum_{\substack{l_1, \dots,l_k, \\ \bigcup_{i=1}^{k} l_i=\mathcal{G}_k}} \hat{\tau}_{m_1,l_1}\dots\hat{\tau}_{m_k, l_k}\;,
\label{Eq::H_eff_linked}
\end{align}
where the sum over $\mathcal{G}_k$ runs over all possible clusters of perturbative order $k$ on an arbitrary lattice and the condition \mbox{$\bigcup_{i=1}^{k} l_i=\mathcal{G}_{k}$} directly reflects that only linked processes have an overall contribution to cluster-additive quantities as stated in the linked-cluster theorem \cite{coester2015optimizing}. As a consequence, we can set up a full-graph decomposition for normal ordering where Eq.~\eqref{Eq::H_eff} is applied to a set of finite, topologically distinct linked  graphs. Usually, the standard approach fails for many different expansion parameters as these serve as another topological attribute \mbox{(the so-called ``link color'')} in the classification of graphs leading to a rapid growth in the number of graphs with increasing perturbative order. For long-range interactions a link color is associated with the coupling strength $\lambda_\delta$ which depends on the distance $\delta$ between the interacting sites. Therefore there are already infinitely many graphs in the first order of perturbation. To overcome this major challenge the use of white graphs \cite{coester2015optimizing} is essential where different colors are ignored in the topological classification and specified only after the calculation on graphs during the embedding in the thermodynamic limit \mbox{(see also next subsection)}. In practice, every link $l_n^{\mathcal{G}}$ of a graph $\mathcal{G}$ is associated with an individual unspecified expansion parameters $\lambda_n^{\mathcal{G}}$. In particular, for models with expansion parameters associated with different coupling flavors $f$ on the same link like the anisotropic XY model, one has to introduce a distinct $\lambda_{n, f}^{\mathcal{G}}$ for each flavor $f\in\{xx, yy\}$. This yields generalized graph contributions as multivariable polynomials which will eventually be substituted with the actual coupling strength and flavor during the embedding procedure of the contributions to the infinite lattice.

\subsection{Embedding}\label{ssec::Embedding}
The normal-ordered effective Hamiltonian in the one quasi-particle (1QP) channel of the anisotropic XY chain with long-range interactions in a transverse field can be expressed as
\begin{equation}
\mathcal{H}_{\text{eff}}^{\text{1QP}} =  \bar{E}_0 + \sum_{j,\delta\ge 0} a_{\delta}(\hat{b}^{\dagger}_{j}\hat{b}_{j+\delta}^{\phantom{\dagger}}+{\rm h.c.} )
\label{Eq::H_eff_1QP}
\end{equation}
in terms of the ground-state energy $\bar{E}_0$ and the 1QP hopping amplitudes $a_{\delta}$. We calculate $\bar{E}_0$ and $a_{\delta}$ in the thermodynamic limit with pCUT as a high-order series in the perturbation parameter $\lambda$ by a graph-embedding procedure. The effective 1QP Hamiltonian can be diagonalized by a Fourier transformation, yielding \mbox{$\tilde{\mathcal{H}}_{\text{eff}}^{\text{1QP}} =  \bar{E}_0 + \sum_{k} \omega(k)\hat{b}^{\dagger}_{k}\hat{b}_{k}^{\phantom{\dagger}}$} where the 1QP dispersion
\begin{equation}
\omega(k)= a_0 + 2\sum_{\delta>0}a_{\delta}\cos(k\delta)
\label{Eq::Dispersion}
\end{equation}
can be read off directly. The elementary excitation gap $\Delta\equiv {\rm min}_k\; \omega(k)$, which we focus on in this work, is located at $k=0$ ($k=\pi$) for ferromagnetic (antiferromagnetic) interactions.
\par
To determine the gap series $\Delta = \sum_{n}c_n\,\lambda^{n}$, the previously calculated white-graph contributions must be embedded into the infinite chain. For long-range interactions every graph can be embedded infinitely many times at any order of perturbation since any pair of spins interacts on the chain. In the following we outline the embedding procedure given in Ref.~\cite{FeySchmidt2D}. 
\par
For the embedding the generic white-graph contributions must be evaluated for each possible realization on the chain by substituting the generic coupling $\lambda_{n,f}^{\mathcal{G}}$ of every link $l_n^{\mathcal{G}}$ of a graph $\mathcal{G}$ with the true coupling depending on the flavor which gives different prefactors depending on the value of $\beta$ and the strength $\lambda|j-i|^{-\alpha}$ where $i$ and $j$ are the sites connected by link $l_n^{\mathcal{G}}$ in the current embedding. Due to the hard-core constraint, two graph vertices cannot be placed on the same lattice site and thus the embedding leads to infinite nested sums which become challenging to evaluate in high orders of perturbation. Therefore, the use of Markov-chain Monte Carlo (MCMC) simulation \cite{FeySchmidt2D, KoziolCylinders} is beneficial for the integration of these sums to provide sufficiently good convergence. In practice, we calculate the coefficients of the gap $\Delta$ directly by
\begin{equation}
c_n=\sum_{N=2}^{n+1}S[f_{N}],
\end{equation}
where $S[\cdot]$ is the Monte Carlo sum over all possible configurations on the chain and the summand $f_{N}$ contains all contributions from graphs with $N$ sites. The fundamental Monte Carlo moves must consist of randomly selecting and moving graph vertices on the lattice. For every embedding each graph contribution is evaluated with the correct coupling strength and added up to the overall contribution. Note, this embedding is less efficient for large values of $\alpha$ in the nearest-neighbor universality.
\par
Using the pCUT method, implemented as a full-graph decomposition, and the MCMC embedding, we have determined the elementary excitation gap as a series expansion up to order 9 in $\lambda$ and in specific cases up to order 10.

\subsection{Extrapolation}\label{ssec::Extrapol}
To determine the quantum-critical regime including critical exponents of second-order phase transitions, we use DlogPad\'e extrapolation for the 1QP gap $\Delta$ beyond the radius of convergence of the bare series. For a detailed description of the following schemes we refer to Ref.~\cite{Guttmann}. 

First, we need to define Pad\'e extrapolants $P[L,M]_{\Delta}$ of the gap series of order $r$ as 
\begin{align}
{\rm P}[L,M]_{\Delta}\equiv\frac{P_L(\lambda)}{Q_M(\lambda)}=\frac{p_0+p_1\lambda+\dots + p_L \lambda^L}{q_0+q_1\lambda+\dots +q_M \lambda^M}\, ,
\label{Eq::Pade}
\end{align}
where $q_0=1$, $p_i, q_i \in \mathbb{R}$ and the degrees $L$, $M$ of $P_{L}(x)$ and $Q_{M}(x)$ with $r\equiv L+M$, i.e., the Taylor expansion of Eq.~\eqref{Eq::Pade} about $\lambda=0$ up to order $r$ must recover the gap series $\Delta$ up to the same order. 
\par 
For the DlogPad\'e extrapolant we define the Pad\'e extrapolant of the logarithmic derivative of the \mbox{gap series $\Delta$} 
\begin{equation}
\mathcal{D}(\lambda)=\frac{d}{d\lambda}\ln(\Delta) \equiv {\rm P}[L,M]_{\mathcal{D}}\, .
\end{equation}
As one degree of the polynomial is lost by differentiating, the degrees must satisfy ${r-1 = L + M}$. Thus the DlogPad\'e extrapolant $d{\rm P}[L,M]_{\Delta}$ is defined as
\begin{equation}
d{\rm P}\left[L,M\right]_{\Delta} \equiv \exp\left(\int_0^\lambda {\rm P}[L,M]_{\mathcal{D}} \,{\rm d}\lambda'\right)\, .
\label{Eq::DlogPade}
\end{equation}
At $\lambda_c$ the Pad\'e extrapolant ${\rm P}[L,M]_{\mathcal{D}}$ exhibits a physical pole which corresponds to a closing of the 1QP gap $\Delta$. We can extract the critical exponent $z\nu$ of the dominant power-law behavior \mbox{$d{\rm P}\left[L,M\right]_{\Delta}\propto |\lambda_c-\lambda|^{z\nu}$} by the residuum of ${\rm P}[L,M]_{\mathcal{D}}$ at the critical point $\lambda=\lambda_c$
\begin{equation}
z\nu := \left. \frac{P_{L}(\lambda)}{\frac{d}{d\lambda}Q_{M}(\lambda)} \right|_{\lambda=\lambda_c}\, ,
\end{equation}
where $\nu$ is the correlation length exponent and $z$ the critical dynamical exponent. Aside from the physical pole it is also possible that spurious poles occur in $0<|\lambda|<|\lambda_c|$ or close to this interval in the complex plane affecting the behavior of the extrapolation at the quantum-critical point $\lambda_c$ and impair the quality of the extrapolation. That is why extrapolations exhibiting spurious poles are called defective and are discarded. We calculate a large set of DlogPad\'e extrapolants with $L+M=r'\le r$, exclude defective extrapolants, and arrange the remaining DlogPad\'es in so-called families with \mbox{$L-M=\text{const}$}. Although individual extrapolations deviate from each other, the quality of the extrapolations increases with the order of perturbation as members of different families but mutual order $r'$ converge. To systematically analyze the gap-closing and associated critical exponents, we take the mean of the highest order extrapolants of different families with more than one member. 
\par
Since the gap displays a power-law behavior near the critical point we assume the gap to be a true physical function $\bar{\Delta}$ of the form
\begin{equation}
\bar{\Delta}(\lambda) \approx \left(1-\frac{\lambda}{\lambda_c} \right)^{z\nu}A(\lambda)
\end{equation}
with $A$ being an analytical function. In case of ferromagnetic interaction, we expect the LRTFIM to show multiplicative logarithmic corrections at the ``lower critical'' $\alpha$ (analogously to the upper critical dimension) in vicinity of the quantum-critical point $\lambda_c$ such that the gap can be expressed as
\begin{equation}
\bar{\Delta}(\lambda) \approx \left(1-\frac{\lambda}{\lambda_c} \right)^{z\nu} \left(\ln\left(1-\frac{\lambda}{\lambda_c} \right)\right)^{p}A(\lambda),
\end{equation}
where $p$ is the associated exponent of the multiplicative logarithmic corrections. Since the extraction of this exponent is very demanding we fix $z\nu$ to the exact mean-field value $1/2$ and the critical point $\lambda_c$ to the value determined by the procedure described above. Now, by defining
\begin{align}
p^{*}(\lambda) &\equiv -\ln(1-\lambda/\lambda_c)[(\lambda_c-\lambda)\mathcal{D}(\lambda)+z\nu] \nonumber\\
&\approx p + \mathcal{O}(\lambda_c-\lambda)
\end{align}
we can extract the multiplicative correction $p$ by employing Pad\'e extrapolations \mbox{$p={\rm P}[L,M]_{p^{*}}|_{\lambda=\lambda_c}$}, again systematically arranging the extrapolants in families and taking the mean of highest order extrapolants.

\section{Isotropic XY chain} 
\label{sec::IXY}
First, we focus on the limiting case of the isotropic XY chain where the tuning parameter $\beta$ is set to zero. As a consequence, terms creating or annihilating two quasi-particles in Eq.~\eqref{Eq::XY_Ham_Op} exactly cancel such that the Hamiltonian solely consists of hopping terms. This leaves 
\begin{align}
\mathcal{H} &=  E_0+\mathcal{Q} - \sum_{i,\delta>0}\lambda_{\delta}(\hat{b}_{i}^{\dagger}\hat{b}_{i+\delta}^{\phantom{\dagger}} + h.c.) \nonumber \\
&= \mathcal{H}_0 + T_0 \equiv \mathcal{H}_{\text{eff}}^{\text{1QP}}\, . 
\label{Eq::H_eff_1QP_isotropic}
\end{align}
In other words, the fully polarized reference state becomes the exact ground state and we recover the effective 1QP Hamiltonian of Eq.~\eqref{Eq::H_eff_1QP} exactly in first order of perturbation which allows us to study the one-particle excitations analytically and to determine the phase transition point including critical exponents. The deduced exponents can be fully understood in terms of a bosonic field-theoretical description. Further, we can investigate the two-particle sector in a quantitative fashion by diagonalizing large finite chains.

\subsection{Gap-closing}
\label{ssec::Q_crit}

For algebraically decaying coupling strengths the 1QP dispersion in Eq.~\eqref{Eq::Dispersion} becomes
\begin{equation}
\omega(k) = 1 - 2\lambda\sum_{\delta > 0} \frac{\cos(k\delta)}{\delta^{\alpha}}\, .
\label{Eq::IXY_Dispersion}
\end{equation}
We determine the quantum-critical point by the closing of the one-particle gap $\Delta$. The associated critical exponents of the second-order quantum phase transition can be extracted by the dominant power-law behavior near criticality of the gap
\begin{equation}
\Delta \propto |\lambda_c-\lambda|^{z\nu}\, ,
\label{Eq::Gap_Exp}
\end{equation}
in the perturbation parameter $\lambda$ with exponent $z\nu$ and the dispersion evaluated at the critical parameter $\lambda_c$
\begin{equation}
\omega|_{\lambda=\lambda_c} \propto |k_c-k|^{z}
\label{Eq::Dispersion_Exp}
\end{equation}
in the quasi-momentum $k$ with the critical dynamical exponent $z$.
\par
\begin{figure}[t]
	\centering
	\includegraphics[width=\columnwidth]{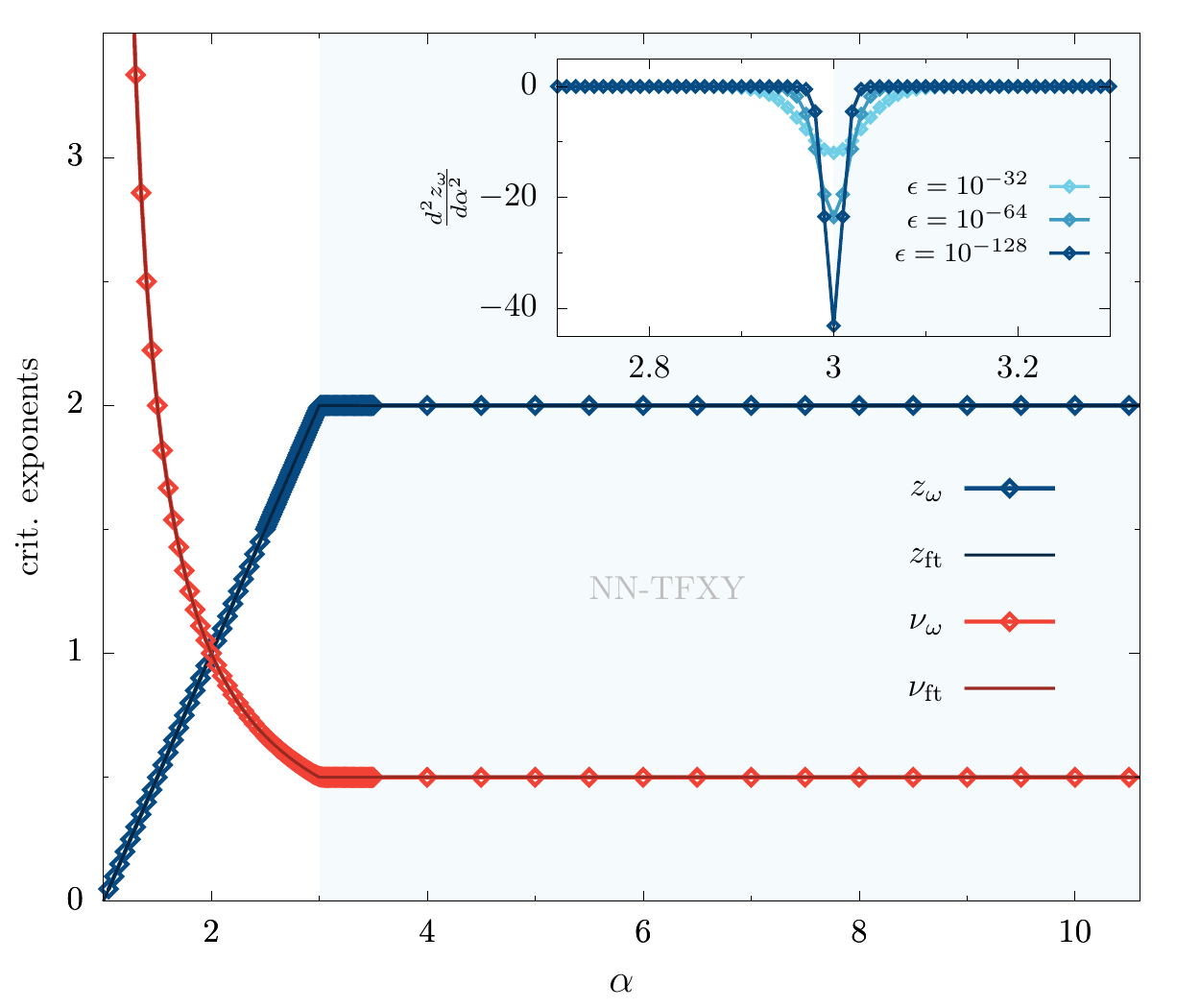}
	\caption{Critical exponents as a function of $\alpha$ for the ferromagnetic transverse-field isotropic XY model. We find the system to be in the nearest-neighbor universality class of the isotropic XY model (NN-TFXY) for $\alpha\ge 3$ where the critical dynamical exponent is $z=2$ (blue lines) and the correlation length critical exponent is $\nu=1/2$ (red lines). In a second regime $\alpha < 3$ the dynamical (correlation length) exponent varies continuously between $0 < z \le 2$ ($1/2 \le \nu < \infty$). The exponents $z_{\text{ft}}$ and $\nu_{\text{ft}}$ are obtained from the field-theoretical treatment while $z_{\omega}$ and $\nu_{\omega}$ are determined by the one-particle dispersion. Both coincide to great accuracy. The second derivative of the critical dynamical exponent $z_{\omega}$ in the inset visualizes that the accuracy of $z_{\omega}$ around $\alpha=3$ strongly depends on the sampling step-size $\epsilon$ of the dispersion around the minimum and therefore the floating point precision. Note, the second derivative approaches a Dirac delta function for $\epsilon\rightarrow 0$ corresponding to a kink in the exponent.}
	\label{Fig::Dyn_Exp}
\end{figure}

In case of ferromagnetic interactions the gap is located at $k=0$ and hence the gap series is given by \mbox{$\Delta=1-2\lambda\zeta(\alpha)$}, where \mbox{$\zeta(\alpha):=\sum_{n=1}^{\infty}n^{-\alpha}$} is the Riemann $\zeta$-function which is convergent for all $\alpha>1$. Coming from the nearest-neighbor limit with $\Delta=1-2\lambda$, a decrease of $\alpha$ shifts the critical point $\lambda_c = 1/2\zeta(\alpha)$ monotonously towards smaller values stabilizing the ordered phase. Since the expression is linear in $\lambda$ the associated critical exponent is always $z\nu=1$ independent of the decaying exponent $\alpha$ of the long-range interaction. Using Eq.~\eqref{Eq::Dispersion_Exp} together with $z\nu=1$, we can determine both critical dynamical and correlation length exponents for any $\alpha>1$ which is depicted in Fig.~\ref{Fig::Dyn_Exp}. 
We find two regimes of different behavior. For $\alpha\ge 3$ both exponents are constant while for $\alpha< 3$ the dynamical exponent (correlation length exponent) continuously varies from $z=2$ to $z=0$ ($\nu=1/2$ to $\nu=\infty$). Note, that no mean-field limit exists due to the specific situation of the isotropic XY model having the fully polarized state as an exact ground-state in the high-field phase. This eigenstate contains no quantum fluctuations and is therefore not sensitive to the variation of $\alpha$. 
\par
Inspired by the quantum-field theory of LRTFIM transitions \cite{Dutta, Defenu2017}, we propose the generic modification of introducing a $k^{\sigma}$-term to the well-studied bosonic action of the isotropic short-range transition \cite{Uzunov1981, Fisher1989, Damle, SachdevBook}, in order to understand the quantum criticality of the long-range XY model from a field-theoretical perspective. Here, we defined $\sigma:=\alpha-d$ where $d$ is the spatial dimension of the system which is one in the case of the XY chain. 
The suggested action then reads as
\begin{align}
S = \frac{1}{2}\int_{k,\omega}(ak^2+bk^{\sigma}+ig\omega+r)|\psi_{k,\omega}|^2+u\int_{x,\tau}|\psi_{x,\tau}|^4
\label{Eq::Action}
\end{align}
with $\psi$ being the complex c-number order-parameter field of the transition, $a,b > 0$ and the real constants $u,g$ and $r$ \footnote{The notation in Eq.~\eqref{Eq::Action} is taken from Ref.~\cite{Fisher1989}.}. From simple power-counting arguments we obtain the critical exponents 
\begin{subequations}
	\begin{eqnarray}
	z &= \begin{cases}2 &\sigma \geq 2 \\ \sigma & \sigma < 2 \end{cases}\\
	\nu &= \begin{cases}\frac{1}{2} &\sigma \geq 2 \\ \frac{1}{\sigma} & \sigma < 2 \end{cases}
	\end{eqnarray}
	\label{Eq::Exps_FT}
\end{subequations}
directly from the Gaussian part of the action in Eq.~\ref{Eq::Action}. These exponents also hold ``presumably''\cite{Fisher1989} for \mbox{$1 \le d \le d_{\text{uc}}$} below the upper-critical dimension $d_{\text{uc}}=2$. Using the arguments provided by Fisher {\it et al.} \cite{Fisher1989} and Uzunov \cite{Uzunov1981} the self-energy vanishes at every order in $u$ and the renormalization of $u$ can be performed in all orders via ladder diagrams. The vanishing self-energy is easily explained by the fact that, in each diagram, every pole in $\omega$ lies in the complex upper half-plane. Hence, the frequency integral can be deformed into the lower half-plane to give zero \cite{Fisher1989}. The fact that the free propagator of the field theory is not changed by the self-energy is a manifestation of the fact that  non-particle conserving fluctuations are absent in the high-field phase. In Fig.~\ref{Fig::Dyn_Exp} we compare the field-theoretical exponents in Eqs.~\eqref{Eq::Exps_FT} with the exponents derived from the 1QP dispersion. The latter coincide with the field-theoretical exponents to an arbitrary accuracy solely depending on the chosen floating point precision.
\par
When restricting to antiferromagnetic interactions the gap at $k=\pi$ can be expressed as $\Delta=1-2\lambda\eta(\alpha)$ with \mbox{$\eta(\alpha):=\sum_{n=1}^{\infty}(-1)^{n-1}n^{-\alpha}$} being the Dirichlet $\eta$-function which is convergent in the entire $\alpha$-regime including ultra long-range interactions in \mbox{$0<\alpha\le 1$}. In contrast to the ferromagnetic case, tuning $\alpha$ towards all-to-all coupling stabilizes the polarized phase shifting the critical point monotonously towards $\lambda_{\rm c}=-\infty$ due to increasing geometric frustration. We find $z=2$ for the critical dynamical exponent over the entire range of $\alpha$-values and, consequently, a constant value $\nu= 1/2$.

\subsection{Two-particle excitations}
\begin{figure}[t]
	\centering
	\includegraphics[width=\columnwidth]{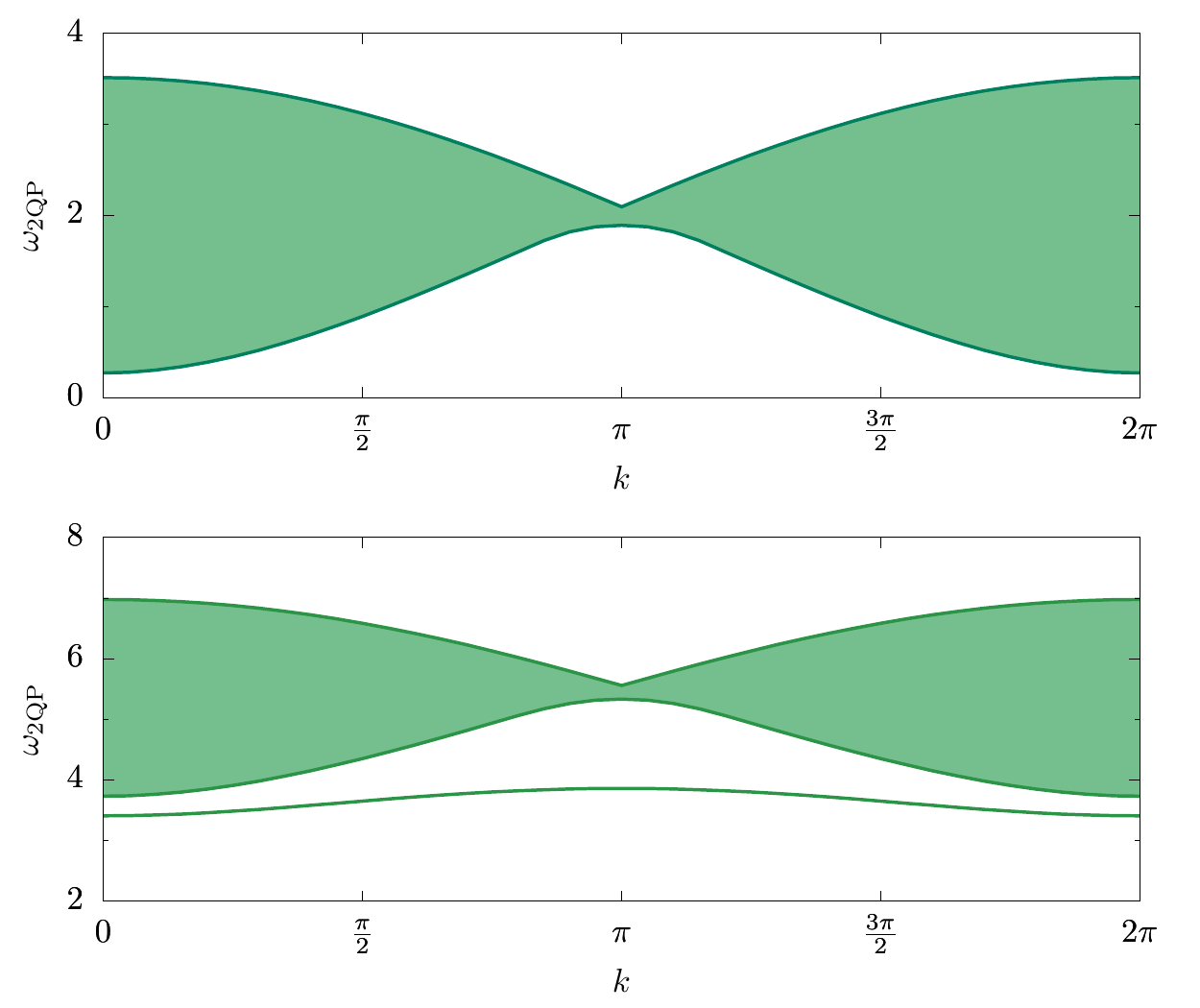}
	\caption{Characteristic continua and bound states in the 2QP sector. Dark green lines highlight the lower and upper continuum edge determined by Eqs.~\eqref{Eq::Dispersion_2QP}. {\it Upper panel}: 2QP continuum for the ferromagnetic transverse-field isotropic  XY model with $\lambda = 0.4$. {\it Lower panel}: Bound state and 2QP continuum of the ferromagnetic transverse-field isotropic XY model in the presence of an additional longitudinal Ising-type coupling with $\kappa=0.4$ and $\lambda =0.4$.}
	\label{Fig::Dispersion_2QP}
\end{figure}

A two-particle excitation is defined as two local spin flips at arbitrary sites $i \neq j$ which we denote as $\ket{\uparrow_i,\uparrow_j}$. In contrast to the previous case the two quasi-particle \mbox{(2QP) Hamiltonian} can not be solved analytically. However, the eigenvalue problem simplifies significantly after Fourier transformation and can be diagonalized numerically with system sizes up to $N=10^5$. In the thermodynamic limit the 2QP eigenenergies form a continuum with the lower band edge $\omega_{\mathrm{lb}}(k)$ and upper band edge $\omega_{\mathrm{ub}}(k)$. Because there is no interaction between the two quasi-particles the lower and upper limits of the continuum are exactly given by the following equations
\begin{subequations}
	\begin{eqnarray}
	\omega_{\mathrm{lb}}(k) &=& \min_q [\omega(k/2+q)+\omega(k/2-q)]\, ,\\
	\omega_{\mathrm{ub}}(k) &=& \max_q [\omega(k/2+q)+\omega(k/2-q)]\, ,
	\end{eqnarray}
	\label{Eq::Dispersion_2QP}
\end{subequations}
only involving the one-particle dispersion. Therefore the resulting two-particle gap located at $k=0$ is exactly two times the one-particle gap. The two-particle continuum is depicted in the upper panel of Fig.~\ref{Fig::Dispersion_2QP}. 
\par
The description of two-particle states as non-interacting combinations of two one-particle states breaks up when establishing an interaction between the two quasi-particles. Exemplary, we add a longitudinal interaction-term with coupling strength $\kappa$ to the isotropic Hamiltonian, \mbox{resulting in}
\begin{align}
\mathcal{H} &=  \frac{1}{2}\sum_{i}  \sigma_{i}^{z} - \sum_{i,\delta>0}\kappa_\delta \, \sigma_{i}^{z}\sigma_{i+\delta}^{z}  \nonumber\\ &- \sum_{i,\delta>0}\frac{\lambda_\delta}{2}\left(\sigma_{i}^{x}\sigma_{i+\delta}^{x} + \sigma_{i}^{y}\sigma_{i+\delta}^{y}\right).
\label{Eq::XY+Z_Ham}
\end{align}
As before, we can calculate the one-particle gap analytically and obtain a similar result just picking up an additional constant term $\kappa\zeta(\alpha)$. However, we again have to solve the 2QP problem numerically. For ferromagnetic interaction with positive $\kappa$, we establish a long-ranged attractive force between the two quasi-particles leading to a bound state which lays below the continuum as shown in the lower panel of Fig.~\ref{Fig::Dispersion_2QP}. We find that the bound state shifts towards the lower continuum edge and eventually decays into the continuum when approaching the critical point as a function of $\lambda$. As a consequence, close to the phase transition, the two-particle gap is again given by twice the one-particle gap equal to the lower band edge at $k=0$. The one- and two-particle gaps therefore close simultaneously. In contrast, for an antiferromagnetic repulsive interaction the system yields an anti-bound state above the continuum (not shown) and therefore the anti-bound state does not play any role in the breakdown of the phase.

\section{Anisotropic XY chain}
\label{sec::AIXY}

In contrast to the previous case, the Hamiltonian in Eq.~\eqref{Eq::XY_Ham_Op} with anisotropic XY interactions is not yet quasi-particle-conserving. Therefore we employ the methods described in Sect.~\ref{sec::Approach} to extract the critical point $\lambda_c$ including critical exponents $z\nu$ associated to the closing of the gap. We discuss the ferromagnetic results by tuning from the limiting case of pure Ising-type interactions \mbox{(see also discussion in~Ref.~\cite{FeySchmidt1D})} towards the isotropic XY limit and then turn to the discussion for antiferromagnetic interactions.

\subsection{Ferromagnetic Case}
\label{subsec::Ferro}
\begin{figure}[t]
	\centering
	\includegraphics[width=\columnwidth]{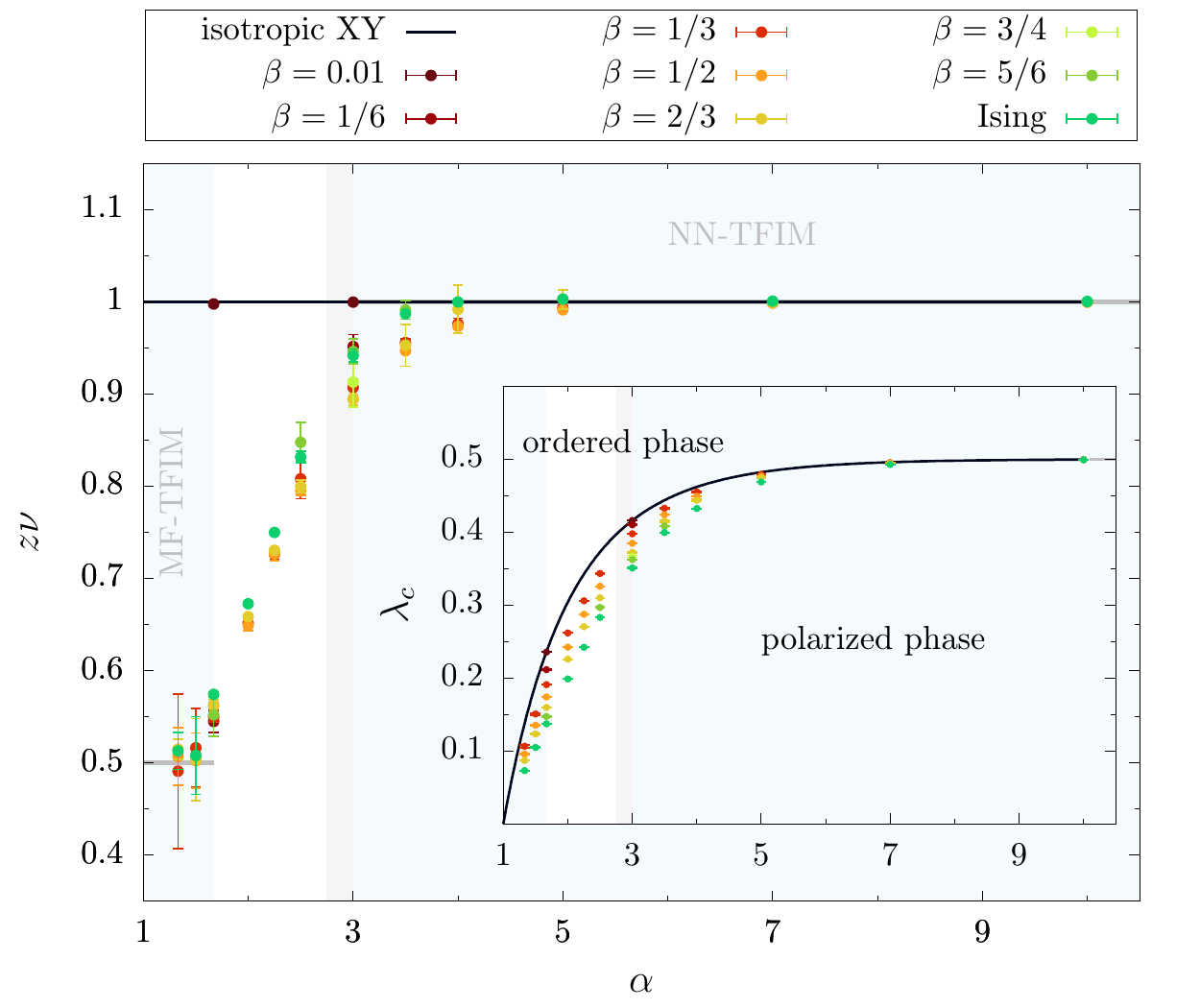}
	\caption{Critical exponents $z\nu$ and critical points $\lambda_c$ (inset) as function of $\alpha$ for the anisotropic XY model in a transverse field evaluated for various values of the interpolation parameter $\beta$ (deep red to green color gradient) with ferromagnetic interactions. The data points are averages from  highest order members of families consisting of non-defective DlogPad\'e extrapolants. When strengthening the long-range coupling the critical point shifts towards zero, i.e., the ordered phase becomes more prominent. For the critical exponent in case of pure Ising-type interactions the system exhibits a nearest-neighbor transverse-field Ising universality (NN-TFIM) for  $\alpha \ge 3$ \mbox{(gray area for $\alpha \ge 2.75$)} with $z\nu=1$, a mean-field domain (MF-TFIM) for $\alpha \le 5/3$ with $z\nu=1/2$ \mbox{(light blue areas and gray lines)} and an intermediate domain with a continuously varying exponent \mbox{(c.f.~Ref.~\cite{Defenu_2020} and references therein)}. The LCE results confirm the presence of these domains for any extrapolation parameter except for $\beta \approx 0$ where the systems jumps to the isotropic behavior as expected (see dark red point at $\alpha=5/3$ for $\beta=0.01$).}
	\label{Fig::Ferro}
\end{figure}

In the limiting case of the LRTFIM at $\beta=1$ the location of the critical point exhibits a similar $\alpha$-dependency as the previously discussed limiting case of isotropic XY interactions at $\beta=0$. However, the ordered phase becomes even more prominent when strengthening the long-range nature of the interactions and in contrast to a constant critical exponent $z\nu=1$ in the entire $\alpha$-range we observe a continuously varying exponent which can be seen in Fig.~\ref{Fig::Ferro}.  
From Refs.~\cite{Fisher1972, Knap, Dutta, Sak, Defenu2017, Defenu_2020} we can identify three distinct regimes. The first domain is for $\alpha\le 5/3$ where the system exhibits mean-field behavior with the associated exponent $z\nu=0.5$. Secondly, for large values of $\alpha$, the system has $z\nu=1$ where the phase transition belongs to the nearest-neighbor transverse-field Ising (NN-TFIM) universality class and thirdly, in the intermediate regime, the critical exponents vary continuously from $z\nu=0.5$ to $1$ in a non-trivial fashion. The exact value of the lower boundary of the nearest-neighbor (NN)-domain is still disputed in literature, while Refs.~\cite{Fisher1972, Dutta, Knap, Suzuki1, Suzuki2, Yamazaki, Picco, Blanchard, Grassberger, Zhangqi} suggest $\alpha=3$ others \cite{Sak, Defenu2017, Gori, Luijten, Angelini, Horita, Paulos, Behan1, Behan2, Behan3} propose $\alpha=2.75$. Like in previous LCE calculations \cite{FeySchmidt1D, FeySchmidt2D} we are able to identify the same three regimes albeit with a larger intermediate realm ($1.5 \lesssim \alpha \lesssim 4$) which can be attributed to the nature of LCEs that cannot resolve a sudden change of the exponent due to the finite order of the bare series and the presence of multiplicative logarithmic corrections at $\alpha=5/3$ that we refer to as ``lower critical'' $\alpha$ in analogy to the upper critical dimension \cite{FeySchmidt1D}.  From perturbative renormalization group calculations and series expansions \cite{Larkin, Brezin, Wegner, Weihong, CoesterHypercube} for the NN-TFIM in three dimensions the multiplicative correction is known to be $p=-1/6$. To extract the correction we calculate the gap for $\alpha=5/3$ up to order 10 in the expansion parameter $\lambda$. As in Refs.~\cite{FeySchmidt1D, FeySchmidt2D}, we can extract the multiplicative logarithmic correction by fixing the exponent $z\nu$ to the exact mean-field value $1/2$ and the critical point $\lambda_c$ to the value previously determined by the DlogPad\'e extrapolants. We find $p=-0.214(8)$ (c.f. Table~\ref{tab:MultiCorrections}) in qualitative agreement with \cite{FeySchmidt1D} again being fairly close to the exact value $-1/6$. \par
Next, we tune the extrapolation parameter $\beta$ from the limit of pure Ising interactions to the limit of isotropic XY interactions. As described before, in the isotropic XY limit all higher-order corrections vanish exactly and the first-order perturbation theory becomes exact. Naively, we expect that uncertainties of the finite perturbative order are reduced and the extrapolation of the LCE becomes increasingly well behaved when tuning to this non-interacting limit. However, simultaneously, the gap as a function of the expansion parameter $\lambda$ flattens such that the critical point exhibits a continuous shift towards larger values of $\lambda$ away from the unperturbed reference point in favor of the polarized phase. Consequently, both effects engage in competition such that families of extrapolants exhibit best convergence at $\beta=0.5$. On that account we observe the largest difference in the extrapolation uncertainties between extrapolations of different values of the interpolation parameter in the intermediate $\alpha$-regime where the critical points differ the most. Further, the critical exponents behave similarly to the pure Ising case, again exhibiting three distinct domains of quantum-critical behavior though with an even larger intermediate regime of continuously varying exponents ($1.5 \lesssim \alpha \lesssim 5$) due to a systematic offset between data points close to the Ising limit $\beta \in \{5/6, 1\}$ and points in the intermediate regime $\beta \in \{1/3, 1/2, 2/3\}$. Since the extrapolants show best convergence at intermediate values of the extrapolation parameter $\beta$ and we naively expect the same behavior as for the LRTFIM, the origin of this systematic offset is unclear. Possibly, as all perturbation orders greater than 1 vanish towards the isotropic XY limit, the relevant physics is shifted to even higher orders such that it becomes harder to resolve the critical exponents and likewise the boundary of the NN-domain. Close to the XY limit the exponents shifts upwards again (c.f. $\alpha=5/3$ or $3$) until the exponent becomes 1 for $\beta=0.01$ in proximity of the multicritical point at the intersection of the Ising and (Ising)$^{2}$ phase transition lines just like in the nearest neighbor case.
\par
Moreover, we extract the multiplicative logarithmic corrections for different values of $\beta>0$ from the gap series in order 10 following the exact same procedure as in the Ising limit. The multiplicative corrections can be found in Table~\ref{tab:MultiCorrections}. Note again, for $\beta=0$ there exists no mean-field limit and therefore no multiplicative corrections are present.
\begin{table}
	\caption{\label{tab:MultiCorrections} Multiplicative logarithmic corrections $p$ at the ``lower critical'' $\alpha$ ($\alpha=5/3$) for different values of the interpolation parameter $\beta$ between the isotropic XY limit ($\beta=0$) and the pure Ising limit ($\beta=1$). From perturbative renormalization group calculations and series expansions \cite{Larkin, Brezin, Wegner, Weihong, CoesterHypercube} we expect $p=-1/6$.}
	\begin{ruledtabular}
		\begin{tabular}{cc}
			Tuning parameter $\beta$ & Multiplicative correction $p$ \\ 
			\hline
			$1/6$ & $-0.240(22)$ \\
			$1/3$ & $-0.2090(32)$ \\
			$1/2$ & $-0.2197(30)$ \\
			$2/3$ & $-0.1957(29)$ \\
			$5/6$ & $-0.1763(24)$ \\
			$1$ & $-0.214(8)$ \\ 			
		\end{tabular}
	\end{ruledtabular}
\end{table}
All values are close to the expected exact value $p=-1/6$. However, as before, the families of biased DlogPad\'es show the best convergence at intermediate values of the tuning parameter $\beta \approx 0.5$.  Despite this observation we find the best estimate at $\beta=5/6$. The fact that all values are systematically larger than $p=-1/6$ can likely be attributed to the sensitivity of $p$ on the position of the critical point and the nature of DlogPad\'e extrapolations that are known to slightly overestimate critical values. Additionally, the relatively small uncertainty in the coefficients of the bare series from the Monte Carlo embedding can spoil the position of the critical point enough to significantly influence the quality of the extracted multiplicative correction. Nevertheless, all values are remarkably consistent with the quantum-field theoretical predictions.
\par

\subsection{Antiferromagnetic Case}
\label{subsec::AF}

\begin{figure}[t]
	\centering
	\includegraphics[width=\columnwidth]{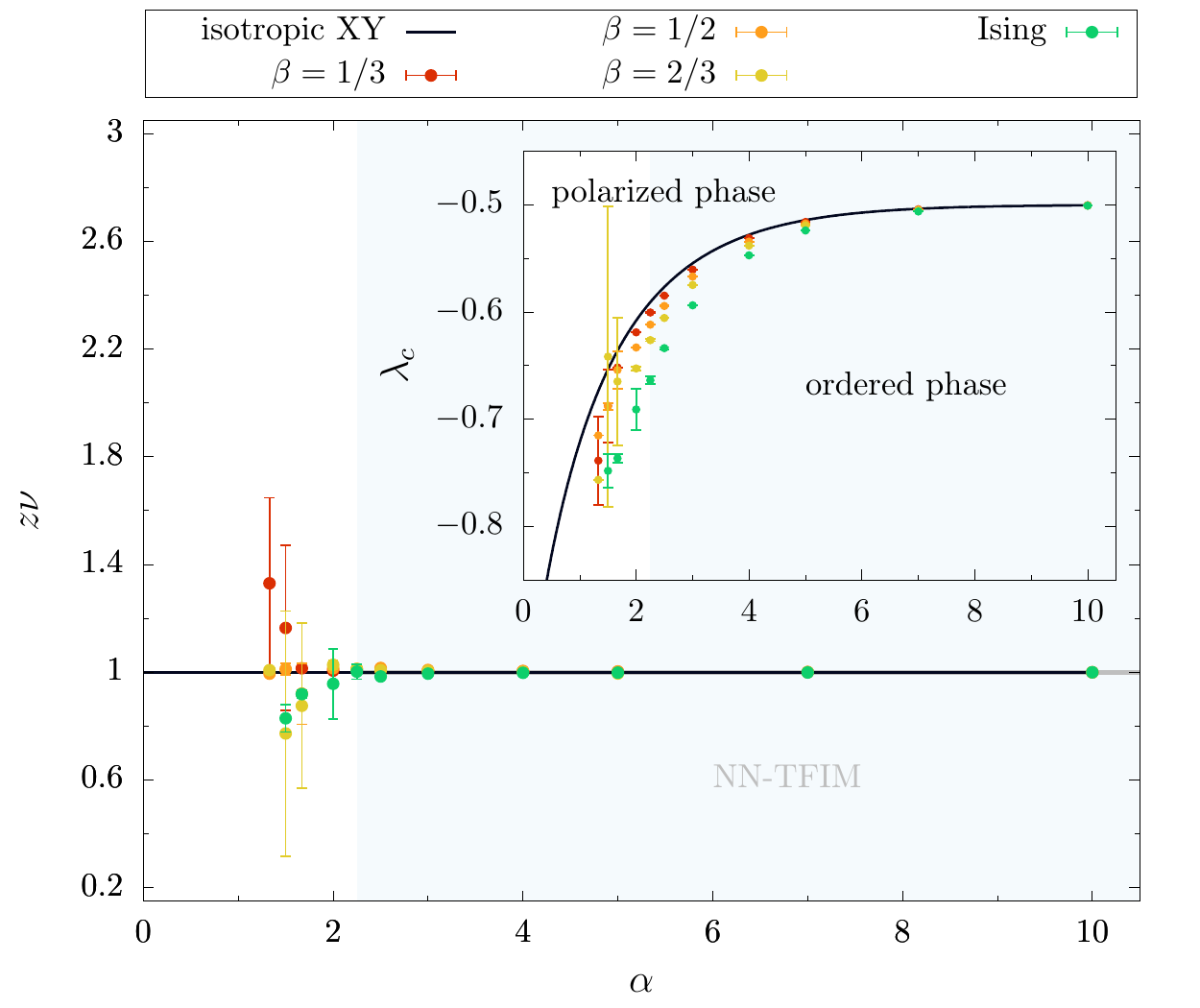}
	\caption{Critical exponents $z\nu$ and critical parameters $\lambda_c$ (inset) as functions of $\alpha$ for the anisotropic XY model in a transverse field evaluated for various values of the interpolation parameter $\beta$ (red to green color gradient) with antiferromagnetic interactions. The data points are averages from  highest order members of families consisting of non-defective DlogPad\'e extrapolants. When strengthening the long-range coupling the critical point shifts towards $-\infty$ such that the polarized phase stabilizes. For pure Ising-type interactions the Refs.~\cite{Koffel, Peter} suggest a NN-domain for $\alpha \ge 9/4$ with $z\nu=1$ (light blue area and gray line) and a second domain for $\alpha < 9/4$ with continuously varying exponent. LCE results, that become increasingly challenging to extract for decreasing $\alpha$, do not indicate a second domain and instead suggest the same NN universality class for $\alpha\le 5/3$.}
	\label{Fig::AF}
\end{figure}

For antiferromagnetic interactions we start again with pure Ising-type coupling. The major difference to ferromagnetic interactions is the presence of geometric frustration for any finite $\alpha$ which makes it difficult to resolve the critical point for values $\alpha \le 2$. This becomes evident in the inset of Fig.~\ref{Fig::AF}. 
We also find that the polarized phase stabilizes when decreasing $\alpha$ even more in the Ising-limit compared to isotropic XY interactions. Renormalization group calculations \cite{Koffel} suggest that the LRTFIM is in the NN-Ising universality class for $\alpha \ge 9/4$. For smaller values of $\alpha$ the physics is less clear. While results from variational matrix product states \cite{Koffel, Peter} suggest continuously varying critical exponents and a breakdown of the area law, DMRG results \cite{Sun} and previous LCEs \cite{FeySchmidt1D, FeySchmidt2D} do not indicate such behavior. Instead we find a constant critical exponent $z\nu=1$ for $\alpha > 2$. For $\alpha \le 2$ less extrapolation families contribute and the critical exponent becomes increasingly challenging to extract. Note, we could not find any valid family for $\alpha=4/3$. Yet, all other exponents are still rather close to $z\nu=1$. 
\par
Again, tuning the extrapolation parameter $\beta$ from the Ising limit to the limit of isotropic XY interactions, we observe a continuous shift of the critical point towards smaller values of the expansion parameter $\lambda$, in contrast to the ferromagnetic case. The critical exponents $z\nu$ for any $\beta$ show the same behavior as in the Ising- and isotropic XY limit. The large deviations for $\beta=2/3$ at $\alpha = 3/2$ and $\alpha=5/3$ can be attributed to the fact that out of four valid families of extrapolants three contribute with a critical exponent very close to 1 while the fourth family contributes with $z\nu\approx 0.4$ but cannot be excluded following the procedure in Subsect.~\ref{ssec::Extrapol}. However, it is still possible that this family might converge to 1 for higher orders. Further, we find increasingly convergent families of extrapolants and usually more families of non-defective DlogPad\'es at smaller $\alpha$-values when tuning the extrapolation parameter towards isotropic XY interactions leading to more realistic estimates for the uncertainties and generally indicating more reliable results. For $\beta=1/3$ the critical exponents are very close to $z\nu=1$ until at least $\alpha=5/3$. The better convergence for smaller $\beta$ is plausible as we tune towards the exactly solvable limit and simultaneously the gap as a function of $\lambda$ steepens such that the critical point is continuously shifted towards smaller values of $\lambda$.

\section{Conclusions}
\label{sec::Conclusion}
In this work we have applied the recently developed high-order linked-cluster expansions for quantum many-body systems involving long-range interactions \cite{FeySchmidt2D} to the high-field polarized phase of the long-range anisotropic XY model in a transverse field. Physically, this allows to tune from the interacting Ising case to the non-interacting isotropic XY model, where first-order perturbation theory becomes exact in the 1QP sector in the high-field phase. As expected, in the antiferromagnetic case, when tuning towards the isotropic limit the high-order series become increasingly well behaved allowing an improved access to the quantum-critical properties of quantum magnets with long-range interactions in the presence of geometrical frustration. On the other hand, in the ferromagnetic case, the quantum-critical properties are best accessible in the intermediate interpolation regime since the elementary excitation gap as a function of $\lambda$ flattens shifting the quantum-critical regime away from the unperturbed high-field limit  when tuning towards the isotropic XY limit. In addition, we were able to estimate multiplicative logarithmic corrections at the ``lower critical'' $\alpha$ remarkably close to the expected value $p=-1/6$ for the entire interpolation range of the model except for the isotropic XY model where no mean-field limit exists. Moreover, in the isotropic limit with ferromagnetic interactions, we determined the critical exponents $z$ and $\nu$ analytically by a bosonic quantum-field theory which are confirmed numerically to an arbitrary precision.

We therefore expect that similar studies in the future for other lattices as well as other types of interactions will yield valuable insights in the physical properties of long-range quantum systems.

\section{Acknowledgments}
We thank S. Fey, M. H\"ormann and A. Langheld for fruitful discussions and gratefully acknowledge the compute resources and support provided by the Erlangen Regional Computing Center (RRZE).

\newpage
\bibliography{bibliography}

\end{document}